# MINIMIZING THE TIME OF SPAM MAIL DETECTION BY RELOCATING FILTERING SYSTEM TO THE SENDER MAIL SERVER

Alireza Nemaney Pour[1] , Raheleh Kholghi[2] and Soheil Behnam Roudsari[2]

[1]Dept. of Software Technology Engineering, Islamic Azad University of Abhar, Iran
`pour@abhariau.ac.ir`
[2]Dept. of IT Engineering, Sharif University of Technology, Kish Island, Iran
`{r.kholghi,Soheil.bh}@gmail.com`

## ABSTRACT

*Unsolicited Bulk Emails (also known as Spam) are undesirable emails sent to massive number of users. Spam emails consume the network resources and cause lots of security uncertainties. As we studied, the location where the spam filter operates in is an important parameter to preserve network resources. Although there are many different methods to block spam emails, most of program developers only intend to block spam emails from being delivered to their clients. In this paper, we will introduce a new and efficient approach to prevent spam emails from being transferred. The result shows that if we focus on developing a filtering method for spams emails in the sender mail server rather than the receiver mail server, we can detect the spam emails in the shortest time consequently to avoid wasting network resources.*

## KEYWORDS

*Anti-spams, Receiver mail server, Sender mail server, Spam Email*

## 1. INTRODUCTION

An e-mail is considered "spam" when a massive number of them are sent to multiple recipients. Spam email is usually used for advertisement or marketing. These unwanted emails cause drawbacks to the recipient, and consume the users' network resources. The disadvantages of spam emails have been addressed in many occasions. In some cases for a single user 9 out of 10 emails are spams that fill his/her inbox. The United States Federal Trade Commission described that 66% of spams have false information somewhere in the message and 18% of spams advertise "Adult" material. According to another report [1] 12% of users spend half hour or more per day dealing with spam emails.

There are several major problems with spam mails. First of all, they are high in volume and fill in mailbox of users. Secondly, there is no correlation between receivers' area of interests and the contents of spam mails. Thirdly, they cost money for ISPs because the bandwidth and the memory of system are wasted. Finally, Spam e-mails cause a lot of security problems because most of them include Trojan, Malwares, and viruses [2].

Many filtering techniques have been developed to control the flow of spam emails. Unfortunately, even with these available techniques, the number of spam emails is growing and the flow has not been controlled completely. The setback is that there is no actual solution because a spammer; an unidentified user with enough knowledge is able to be familiar with the logic of the filtering mechanisms. As a result, bypassing the filter and sending the spam emails seems not to be a difficult task for such spammers. In such cases, the spam emails are not detected and are considered as legitimate ones.

There are studies regarding spam email filtering [3-11]. The common issue with the usage of all of these techniques is that the filtering systems are set up in the receiver mail server, consequently, causing network load and wasting network resources. To preserve network resources such as bandwidth and memory, and to reduce network load, this paper proposes to locate spam email filtering in the sender mail server rather than the receiver mail server. Moreover, this paper by experimental results shows that this novel approach works more efficiently compared with the previously proposed approaches.

This paper is organized as follows. In section 2, the related work to the subject will be highlighted. The Overview of email system and its operation are described in section 3. Our proposal and the experiment results are presented in section 4 and 5 respectively. Finally, the conclusion is shown in section 6.

## 2. RELATED WORK

As stated before, there are many filtering techniques to stop the flow of spam emails to mail boxes [3-12]. Figure 1 simply illustrates the classification of spam email filtering techniques. The classification includes list-based filtering [3-7], static algorithm [8-10], and IP-based filtering [11]. The list-based filtering is classified into three categories; Blacklist [3], Whitelist [4, 5], and Greylist [6, 7]. Static algorithm is classified into content-based [8, 9], and the rule-based [10] filtering. Finally, IP-based filtering consists of revers-lookup [11].

In the Blacklist filtering [3], the IP address and the domain name of the sender server is stored in a list called Blacklist and the emails from that IP address and domain are blocked. Then, based on the policy of the receiver side, the emails from the Blacklisted IP addresses are deleted or sent to spam folder. Conversely, there are some limitations for the Blacklist filtering. First, since the spammer uses several IP addresses with a variety of domain names, updating these lists is a difficult task for the client. Consequently, updating the Blacklist regularly is costly. Second, Blacklist filtering may result in identification of an email as false negative because of minimal control in this methodology.

On the other side of the Blacklist, is the Whitelist filtering [4, 5]. In this technique, any user stores his/her email contacts in a list called the Whitelist. Therefore, any received email with the correspondent address from this list is accepted, and all other addresses out of this list are considered uncertain. In this technique, also there are certain obstacles. The obvious one is that, since the sender is unidentified and unpredictable, it is difficult to insert all possible sender addresses in this list. Similar to the Blacklist, the Whitelist filtering needs to be updated regularly; which is a costly task for the user. Another major issue is that if the email address of a spammer is added in the Whitelist of an email client once, this will provide access to all of the addresses in the Whitelist of that specific client without any boundaries or limits. As a result, this will ensure the spammer more reachable email addresses.

In Greylist filtering [6], a different approach is practiced. This technique can be set either on the mail server of the receiver or/and on the personal receiver anti-spam application. At first step, all received emails are rejected. Because of this policy, spammers do not try to resend the rejected email since it is time consuming for them. Instead, the spammers prefer to search for another email address without Greylist filtering. Moreover, from the behavior of spammer viewpoint, the Greylist by itself provides a usable contact list of the mail servers using this filtering technique, subsequently; the spammer avoids sending more messages to those servers after the first rejection because spammer can recognize what the receiver mail server filtering structure is, and how it operates. Consequently, the spammer will update his techniques in order to bypass Greylist filtering. Finally, a major weakness observed in this technique is that there is a possibility that legitimate messages may be lost [6, 7].

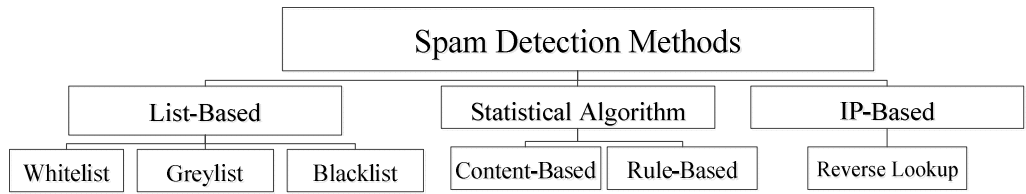

Figure. 1 Classification of Spam Detection Methods

Content-based filtering [8, 9] is another filtering technique that uses machine learning criteria. In order to have the satisfactory results, the administrator of the mail server needs to train the filters to perform their functions. This filtering starts to work based on some predefined words after the email is received entirely. These particular words are collected by statistical reports based on the words and phrases gathered from the spam emails. Rule based filtering [10] is similar to content-based one with some differences. This technique works through some certain rules and regulations. By these rules the filter decides to pass or to block the received email.

The major problem with the content and rule based filtering is that, the rules and the words are verified by the programmer. This leads to variable restrictions. First, the databases and the policies need to be updated at regular basis. Second, as all spammers are aware of these filters, and their functionality, they will try to deliver their messages using additional characters to legitimize their emails. Finally, these techniques work after the body of the email is completely received by the mail server which increases the time for checking whether the email message is spam or not.

In reverse lookup, also known as a reverse DNS (Domain Name System) lookup, the host is associated with a given IP (Internet Protocol) address. By using this routine, the receiver can confirm the identity of the domain name of the sender. This technique is not effective for the mobile users and the users with invalid IP address [11].

Authors in [12] introduce a new procedure based on the spammer behavior. Commonly, a spammer sends an email(s) to huge number of users. In this filtering, the administrator sets a counter on mail server to limit the number of the emails which its clients wish to send. This counter-based filtering provides time saving because the mail server can decide whether a mail is spam or not before the message is completely received. But its restriction is that the legitimate emails may not pass the counter filter.

As spammers become more dominant, the number of anti-spam methodologies and software are growing correspondingly. The problem is that even with the most accurate anti-spam techniques, we lose lots of network resources such as time and bandwidth because these techniques are set on the receiver server side. In most of web based email services such as Hotmail, AOL, and Yahoo, filtering emails start after they are fully received by the receiver mail servers.

In this paper we propose shifting the location of the filtering system from the receive mail server to the sender mail server to achieve efficient results. For this purpose, we define four scenarios and evaluate the results with two anti-spam software, DSPAM[1], and TREC[2]. These software systems are open source programs, and include all filtering techniques stated above.

---

[1] DSPAM: http://dspam.nuclearelephant.com/
[2] TREC: Text REtrieval Conference. http://trec.nist.gov/

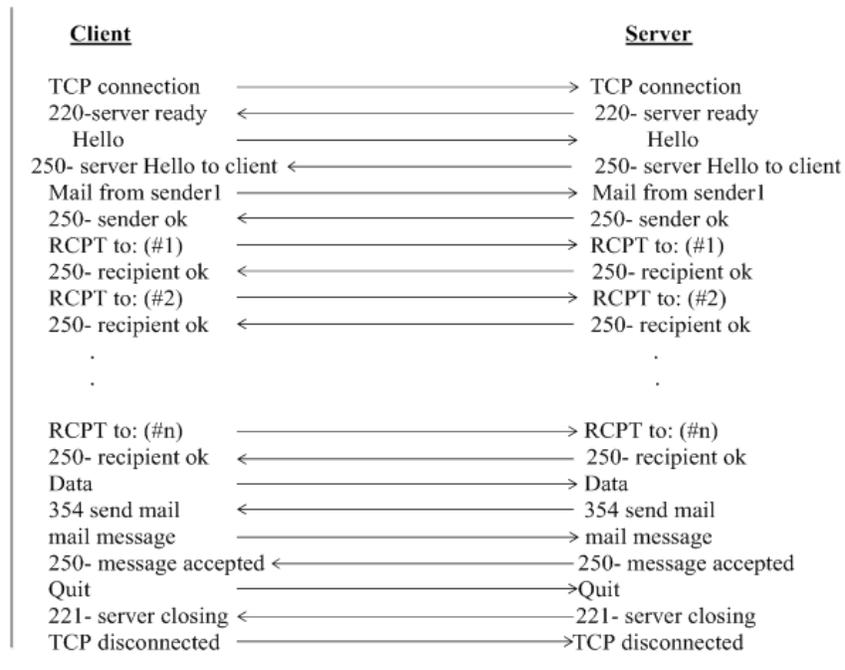

Figure. 2 SMTP Transaction Commands

## 3. OVERVIEW OF EMAIL SYSTEM

In this section, a brief explanation of email protocol and the process of filtering will be elaborated. Simple Mail Transfer Protocol (SMTP) is the first protocol which transfers the emails by some commands. Figure 2 illustrates SMTP commands. First, TCP/IP (Transmission Control protocol and Internet Protocol) connection starts between sender and the associated mail server. Following that, the SMTP commands begin with a Hello message and announcing the acceptance of the session between the client and the server. This process ends when the message is accepted by the mail server. TCP connection disconnects if there is no more message from the client to the mail server.

When the email is delivered by the server, the filtering phase is started. Based on the server filtering policy, Blacklist and Whitelist filtering is stared to examine if the email is a spam or a valid one. If the email is recognized as a valid one, it is sent to receiver's inbox otherwise the email is blocked or transferred to the spam folder. When a Greylist filtering is used in relevant mail server, the email is rejected for the first time. Afterward the body of the email is tested with content-based and rule-based filters according to the standards of the administrator.

## 4. STRUCTURE OF OUR PROPOSAL

This section describes our proposal. The main purpose of this approach is to detect spam emails in the shortest time and consequently to avoid wasting network resources by shifting the location of the filtering system from the receiver mail server to sender mail server (Figure 3). In this way, all emails are screened and checked before being permitted to proceed to the receiver mail server. Figure 4 illustrates the structure of sending an email with necessary filtering steps. The procedure consists of four steps, IP validity check for sender and receiver, list-based filtering, and statistical algorithm.

① First, the validity of IP address of the sender is checked by its mail server. When the client is invalid, the connection is terminated. Otherwise, the mail is preceded to next step.
② Second, the validity of the email address of the receiver is checked by the DNS server. When the email address of the receiver is not available, or wrong, the mail server sends back a failure message to the sender, and the connection is terminated. Otherwise, the mail is preceded to list-based filtering.
③ Third, the email should be checked by the list-based filtering such as Whitelist, Blacklist, and Greylist. By passing these filtering the email is preceded to next step. When an email cannot pass one of these filtering, the connection is terminated.
④ Forth, the statistical filtering such as, content-base and rule-based filtering is started. Emails that cannot be verified as valid ones are not sent.
⑤ Finally, the valid emails are sent to the receiver mail server.

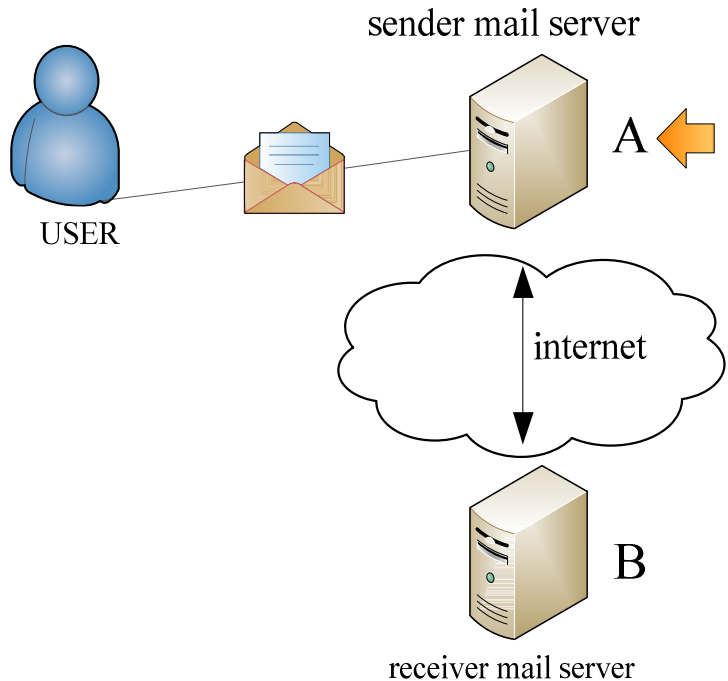

Figure. 3 Shifting the location of filtering system to sender mail server

International Journal of Network Security & Its Applications (IJNSA), Vol.4, No.2, March 2012

```
Program for spam detection
var pass= 1
Begin
If sender IP address = 'valid' then        {sender Authentication}
     goto receiver identification
else
     halt
if receiver IP address = 'valid' then      {receiver Authentication}
     goto list-based filtering
else
     'Failure message'
     halt
if email = pass then         {List-based filter checking}
     goto statistical filtering
else
     halt
if email= pass then
     send to the relevant server
     forwarding
else
     halt
```

Figure. 4 An algorithm of filtering in sender mail server

Our proposal is a new approach to prevent spam emails. Its novelty is that we introduce the filtering methods in the sender mail server rather than the receiver mail server. By this approach, invalid emails are not transferred to the receiver mail server because of filtering in the sender mail server. Consequently, the network resources, such as bandwidth, time, and memory is preserved. In the next section, we will illustrate the efficiency of this approach by fortifying it with experimental results.

## 5. EXPERIMENT RESULTS

This section describes experiment results. We analyse our proposed model and compare the performance of filtering system when is set up on different locations. Figure 5 illustrates the observation model assuming that the spammer is going to send *1000* emails through the mail server *A*. Four scenarios are defined based on changing the location of filtering system as follows:

(1) The spammer is going to send *1000* emails through the mail server *A* to the mail server *B*, and *B* checks the emails (Figure 5 scenario 1).

(2) The spammer is going to send *1000* emails through the mail server *A* to different mail servers such as *B, C, D*, and they check the received emails respectively (Figure 5 scenario 2)

(3) The spammer is going to send *1000* emails through the mail server *A* to mail server *B*, and *A* checks the emails (Figure 5 scenario 3).

(4) The spammer is going to send *1000* emails through the mail server *A* to different mail servers such as *B, C, D*, and the mail server *A* checks the emails (Figure 5 scenario 4).

58

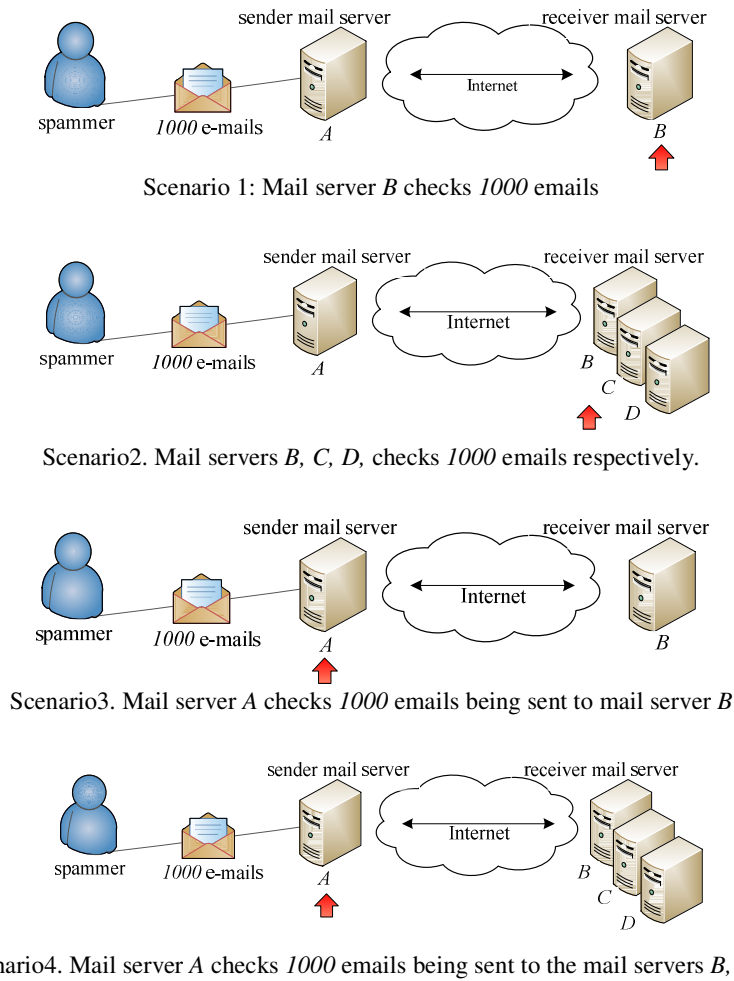

Scenario 1: Mail server *B* checks *1000* emails

Scenario2. Mail servers *B, C, D,* checks *1000* emails respectively.

Scenario3. Mail server *A* checks *1000* emails being sent to mail server *B*

Scenario4. Mail server *A* checks *1000* emails being sent to the mail servers *B, C, D*

Figure. 5 Observation model assuming that the spammer sends *1000* emails through the mail server *A*

For our experiment, we used the results that concluded from various sources [13]. Figure 6 illustrates the result of the performance of anti-spam software running on different ISPs compared with two representative open source anti-spams software. This Figure shows the time required to process *1000* emails in the receiver mail server. The required time to process emails at ISPs such as Hotmail, AOL and Microsoft are, *0.1*, *0.09*, and *0.1* second respectively. On the other hand, the results of our experiment with two representative anti-spam software such as TREC and DSPAM shows that the required time to process *1000* emails are *200* and *250* seconds. This result shows that the performance of email filtering in ISPs is better compared with open source software. The factors that directly affect the performance of filtering discrepancies are based on the size of email, processor power, and several others. Later, we will use the outcomes of this experimentation for the four scenarios explained above.

As experimental results illustrated in Figure 7, we fulfilled our four scenarios. Starting from scenario 1, the mail server *B* checks all *1000* emails one by one. For this purpose, the mail server *B* puts each single email in its specific memory based on each IP address in the email. As a result, the server consumes more time for the same email. The result of this scenario for each anti-spam system is shown with the rightmost bar in Figure 7.

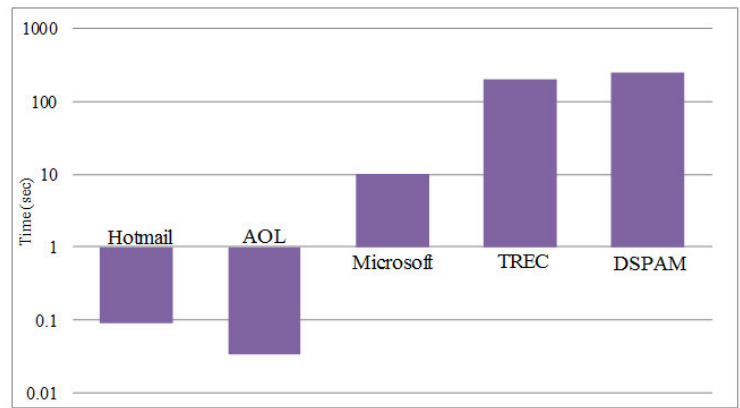

Figure. 6 The performance comparison of filtering between ISPs and open source software in receiver mail server

In scenario 2 the receiver mail servers, *B*, *C* and *D* check all the received emails. In order to send each email, the mail server *A* is required to establish a session with associate mail servers. For this purpose, the receiver mail servers, *B*, *C*, and *D* should accept the sessions for each IP address. The result of this scenario is similar to the first scenario because we consider the total process time for scenario 2. Results of this scenario are shown with blue bar in Figure 7.

In the third scenario the spam filtering is performed in the sender mail server. When the spammer attempts to send *1000* emails to different clients, the anti-spam software starts to process filtering on the email just once. If the email is recognized as trustable email, it is sent to *B*. Otherwise this email will be deleted. The advantage of filtering in the sender mail server is that the network resources between the servers are preserved. The results are displayed with green bars in Figure 7.

In the last scenario, the sender sends *1000* emails to several recipients on different mail servers.

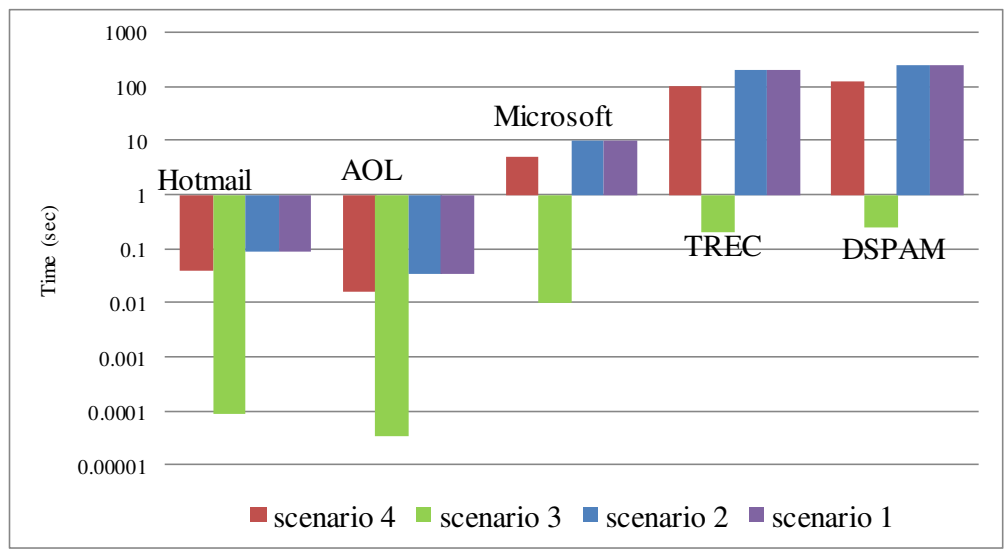

Figure. 7 The performance comparison of spam mail filtering between sender and receiver mail servers

This time the mail server *A* should check the emails. Before any action for filtering, the mail server A needs to establish sessions with the other mail servers. For this reason, the process time is longer than the time required in scenario 3. The anti-spam processing in the sender mail server uses less time when compared the same procedure performed on the receiver mail server. Results are demonstrated with red bars.

The outcome of these assessments, as shown in Figure. 7, imply that scenarios 3 and 4 have better performance compared with scenarios 1 and 2 under all conditions. It is because the filtering process is performed in the sender mail server. Moreover, the performance of scenario 4 is not satisfactory compared with scenario 3 because the sender mail server needs to check the domain of each receiver mail server. In this case, the performance level falls, however still it is considerably better than the cases where the filtering is checked in the receiver mail server (scenarios 1 and 2). Although the results for TREC and DSPAM are based on our experiments, the results for ISPs such as Hotmail, AOL and Microsoft have been calculated logically based on our prediction extracted from Figure. 6. On the other hand, scenario 3 indicates that when the filtering system is located in the sender mail server, the processed time becomes *n* times less than the time when the filtering system is in the receiver mail server when *n* indicates the number of processed emails.

Spam mail detection is a challenging work against human mind because spammers try to find new ways to bypass filtering systems. Therefore, it seems that it is a difficult task to read the spammers' mind and to find all the possible tricks that they might develop. To overcome this problem, we suggest developing methods to provide high performance in the shortest time. Spam mail filtering in the sender mail server (our proposal) is one of those methods compared with the filtering in the receiver mail server.

## 6. CONCLUSION

In this paper, we have proposed an efficient approach for spam email detection. Our approach proposes to shift the location of spam email filtering system from receiver mail server to sender mail server. The purpose of this novel idea is to detect spam emails in the shortest time and consequently to prevent wasting the network resources from misusage of spammers. In addition, by experimental results we proved that our idea is efficient because just the resources in the sender side are accessed. This implies that if an email is identified as spam one, the receiver's bandwidth and memory is preserved which will assure a better performance. Finally, by locating the filtering system in the sender mail server; the processed time becomes *n* times less than the time when the filtering system is in the receiver mail server when *n* indicates the number of processed emails.

This time the mail server *A* should check the emails. Before any action for filtering, the mail server A needs to establish sessions with the other mail servers. For this reason, the process time is longer than the time required in scenario 3. The anti-spam processing in the sender mail server uses less time when compared the same procedure performed on the receiver mail server. Results are demonstrated with red bars.

The outcome of these assessments, as shown in Figure. 7, imply that scenarios 3 and 4 have better performance compared with scenarios 1 and 2 under all conditions. It is because the filtering process is performed in the sender mail server. Moreover, the performance of scenario 4 is not satisfactory compared with scenario 3 because the sender mail server needs to check the domain of each receiver mail server. In this case, the performance level falls, however still it is considerably better than the cases where the filtering is checked in the receiver mail server (scenarios 1 and 2). Although the results for TREC and DSPAM are based on our experiments, the results for ISPs such as Hotmail, AOL and Microsoft have been calculated logically based on our prediction extracted from Figure. 6. On the other hand, scenario 3 indicates that when the filtering system is located in the sender mail server, the processed time becomes *n* times less than the time when the filtering system is in the receiver mail server when *n* indicates the number of processed emails.

Spam mail detection is a challenging work against human mind because spammers try to find new ways to bypass filtering systems. Therefore, it seems that it is a difficult task to read the spammers' mind and to find all the possible tricks that they might develop. To overcome this problem, we suggest developing methods to provide high performance in the shortest time. Spam mail filtering in the sender mail server (our proposal) is one of those methods compared with the filtering in the receiver mail server.

## 6. CONCLUSION

In this paper, we have proposed an efficient approach for spam email detection. Our approach proposes to shift the location of spam email filtering system from receiver mail server to sender mail server. The purpose of this novel idea is to detect spam emails in the shortest time and consequently to prevent wasting the network resources from misusage of spammers. In addition, by experimental results we proved that our idea is efficient because just the resources in the sender side are accessed. This implies that if an email is identified as spam one, the receiver's bandwidth and memory is preserved which will assure a better performance. Finally, by locating the filtering system in the sender mail server; the processed time becomes *n* times less than the time when the filtering system is in the receiver mail server when *n* indicates the number of processed emails.

International Journal of Network Security & Its Applications (IJNSA), Vol.4, No.2, March 2012

**Authors**

**Alireza Nemaney Pour**[1] has obtained his B.S degree in Computer Science from Sanno University, Japan, M.S in Computer Science from Japan Advanced Institute of Science And Technology, Japan, and Ph.D. degree in Information Network Science from Graduate School of Information Systems, the University of Electro-Communications, Japan.
He is currently a faculty member of Islamic Azad University of Abhar in Iran. In addition, He is a technical advisor of J-Tech Corporation in Japan. His research interests include Network Security, Group Communication Security, Protocol Security, Information Leakage, Spam Mail Prevention, Web Spam Detection, and Cryptography.

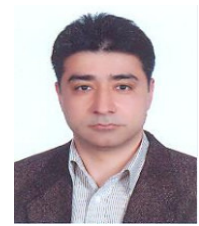

**Raheleh Kholghi**[2]  has received her B.Sc. from Azad University in software engineering and M.Sc. degree in Information Technology Engineering from Sharif University of Technology, International Campus, Iran in 2011. She works currently in Telecommunication Kish Company (TKC) as an IT engineer. Her research area is about network and its application security.

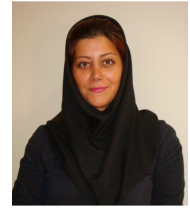

**Soheil Behnam Roudsari**[2] has received his B.Sc. in Information Technology Engineering from Sharif University of Technology, International Campus, Iran. He is currently a M.Sc. student in Engineering of Computer Systems, Politecnico Di Milano University, Italy. His area of research interest is Network Security.

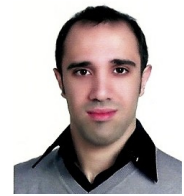